\documentclass[12pt,a4paper]{article}

\usepackage{graphicx}                
\usepackage{microtype}
\PassOptionsToPackage{warn}{textcomp}
\usepackage{textcomp}                
\usepackage{mathptmx}                
\usepackage{times}                   
       
\usepackage{cite}                    
\usepackage{tabu}                    
\usepackage{booktabs}                
\usepackage{amsmath}
\usepackage[ruled,linesnumbered]{algorithm2e}

\title{Distance Adjustment of a Graph Drawing Stress Model}

\author{
  Yosuke Onoue\\
  Nihon University \\
  onoue.yousuke@nihon-u.ac.jp
}
\date{}

\begin{document}
\maketitle

\begin{abstract}
  Stress models are a promising approach for graph drawing.
  They minimize the weighted sum of the squared errors of the Euclidean and desired distances for each node pair.
  The desired distance typically uses the graph-theoretic distances obtained from the all-node pair shortest path problem.
  In a minimized stress function, the obtained coordinates are affected by the non-Euclidean property and the high-dimensionality of the graph-theoretic distance matrix.
  Therefore, the graph-theoretic distances used in stress models may not necessarily be the best metric for determining the node coordinates.
  In this study, we propose two different methods of adjusting the graph-theoretical distance matrix to a distance matrix suitable for graph drawing while preserving its structure.
  The first method is the application of eigenvalue decomposition to the inner product matrix obtained from the distance matrix and the obtainment of a new distance matrix by setting some eigenvalues with small absolute values to zero.
  The second approach is the usage of a stress model modified by adding a term that minimizes the Frobenius norm between the adjusted and original distance matrices.
  We perform computational experiments using several benchmark graphs to demonstrate that the proposed method improves some quality metrics, including the node resolution and the Gabriel graph property, when compared to conventional stress models.
\end{abstract}

\begin{figure}[t]
  \centering
  \includegraphics[width=\linewidth]{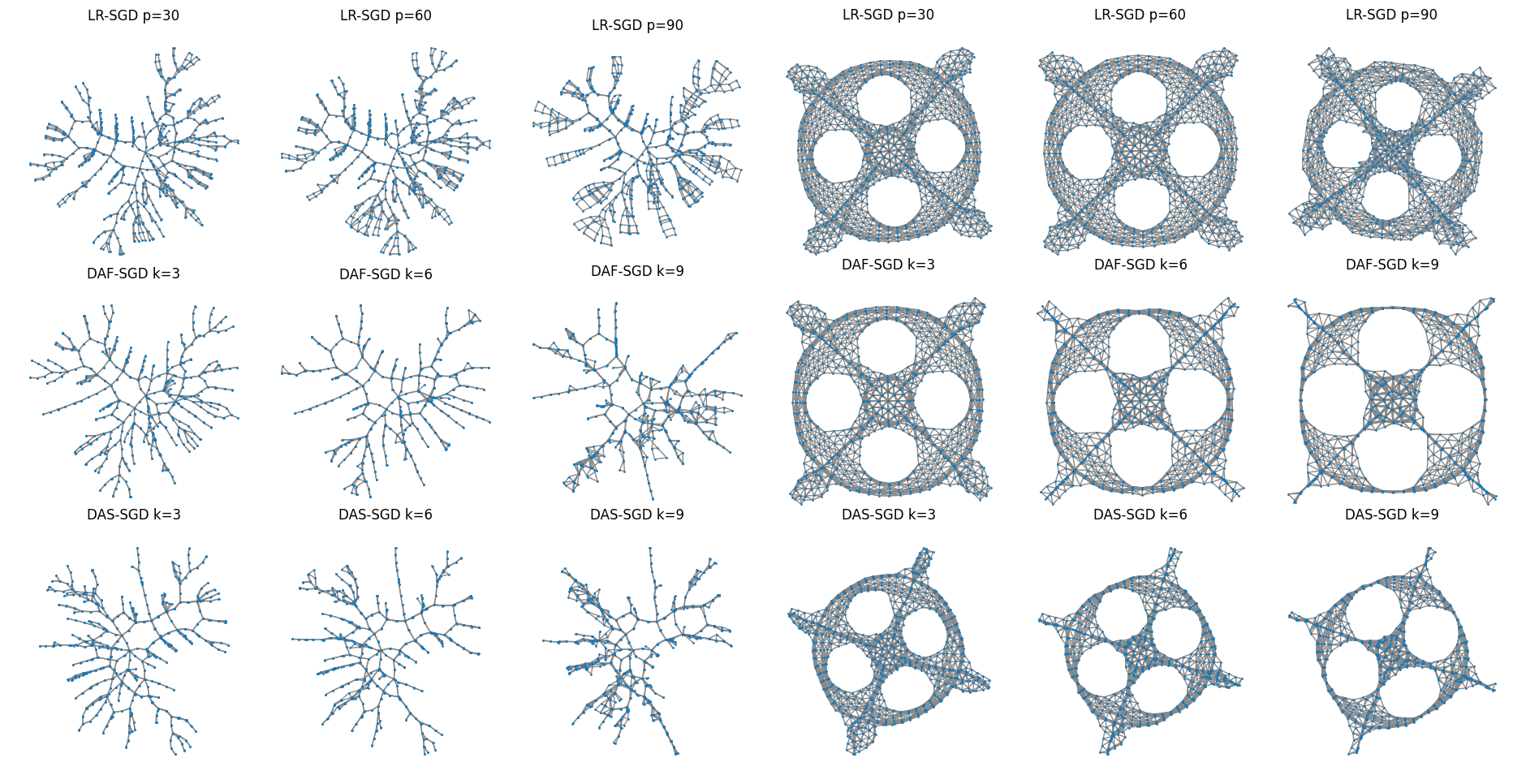}
  \caption{%
    Drawing results of qh882 and dwt\_1005 graphs using our propoed methods Low-rank SGD (LR-SGD), Distance-adjusted FullSGD (DAF-SGD), and Distance Adjuted SparseSGD (DAS-SGD).
    $p$ in LR-SGD and $k$ in DAF-SGD an DAS-SGD are parameters expressing the strength of distance adjustment.
    The stronger the ditance adjustment, the more the detailed structure of the graph is simplified and the overall rough structure is emphasized.
  }
  \label{fig:teaser}
\end{figure}

\section{Introduction}

Graph drawing computes low-dimensional (usually 2 or 3)node coordinates from an input graph and is widely studied in the visualization field\cite{tamassia2013handbook}.
Stress models \cite{kamada1989algorithm,gansner2005graph,zheng2018graph} are a promising approach in graph drawing.
The input graph is represented as $G = (V, E)$, where $V$ is the node set, and $E \subseteq V \times V$ is the edge set.
In this paper, the graph is assumed to be a simple undirected connected graph.
The stress model for obtaining $m$-dimensional coordinates minimizes the following stress function:
\begin{equation}
  S(X) = \sum_{i < j} w_{ij} (|X_i - X_j| - d_{ij})^2
\end{equation}
where $X = (X_1, X_2, \cdots, X_{|V|})^T$ and $X_i = (x_{i1}, x_{i2}, \cdots, x_{im})^T$ are the $m$-dimensional coordinates of node $i$; $d_{ij}$ is the desired distance between nodes $i$ and $j$; and $w_{ij}$ is the weight usually set to $d_{ij}^{-2}$.
The distance matrix typically uses the graph-theoretic distances obtained from the all-node pair shortest path problem.
The stress model minimizes the potential energy through the elastic force of the system, in which all nodes are connected by springs.

\begin{figure}[h]
  \centering
  \includegraphics[width=8cm]{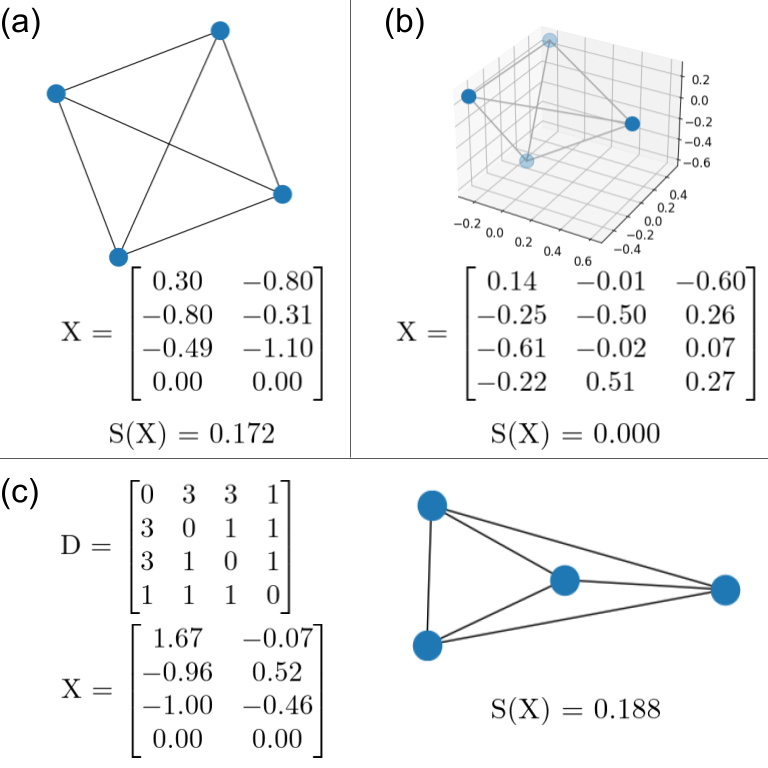}
  \caption{
    Examples of distance relations and complete graph drawing of degree 4.
    In Case (a), the nodes that satisfy the given distance relationship cannot be arranged in a two-dimensional space.
    In (b), the distance relationship similar to that in (a) is possible in a three-dimensional space.
    In (c), the distance relationship cannot be satisfied, even in a high-dimensional space.
  }
  \label{fig:minimum_stress}
\end{figure}

The lower bound of the stress function is 0, but in what cases can it be achieved?
Whether or not the stress function will have a minimum value of 0 depends on the distance matrix $D$ defined as follows:
\begin{equation}
  D = \left[ d_{ij} \right]_{|V| \times |V|, (1 \leq i \leq |V|, 1 \leq j \leq |V|)}
\end{equation}
For it to be achieved, the inner product matrix calculated from the distance matrix must be a positive semidefinite and must have no greater than $m$ positive eigenvalues.
However, it is generally not achievable for the two following reasons in the coordinates composing the distance matrix:
(1) the inner product matrix becomes an indefinite value matrix because the coordinates cannot be expressed in the Euclidean space, and
(2) the inner product matrix has more than m nonzero eigenvalues because the coordinates cannot be represented in m dimensions.
Figure \ref{fig:minimum_stress} shows examples of a complete graph drawing of degree 4.
When the stress function is minimized with these properties included, the obtained coordinates are affected by the non-Euclidean property and the high-dimensionality of the distance matrix.
Therefore, the graph-theoretic distances used in stress models may not necessarily be the best metric for determining the node coordinates.

Our research question begins with how the proper adjustment of the distance matrix affects the graph drawing results.
We propose herein two different methods of adjusting the graph-theoretical distance matrix to a distance matrix suitable for graph drawing while preserving its structure.
The first method is the application of eigenvalue decomposition to the inner product matrix obtained from the distance matrix and the obtainment of a new distance matrix by setting some eigenvalues with small absolute values to zero.
The second approach is the usage of a stress model modified by adding a term that minimizes the Frobenius norm between the adjusted and original distance matrices.
Both methods are thought to work with minor modifications to existing algorithms using stress models, such as the Kamada--Kawai algorithm \cite{kamada1989algorithm}, stress majorization \cite{gansner2005graph}, and stochastic gradient descent (SGD) \cite{zheng2018graph}.
In this study, we specifically integrate the proposed distance adjustment method into SGD.
The SGD variants incorporating the former and latter methods are low-rank SGD (LR-SGD) and DA-SGD, respectively.
Apart from stress, we employ eight quality metrics to assess the effect of distance adjustment and perform computer experiments on several well-known benchmark graphs.
The results show that an appropriate distance adjustment improves some quality metrics, including the node resolution and the Gabriel graph property, when compared to conventional stress models.

The main contributions of this study are as follows:
\begin{enumerate}
  \item We focus on the distance matrix widely used in stress models and clarify that a proper adjustment of the distance matrix appropriately affects the drawing results.
  \item We propose two specific methods for the distance matrix adjustment:
        (1) low-rank approximation of the distance matrix: a method based on the eigenvalue decomposition of a distance matrix; and
        (2) distance-adjusted stress model: a method using a modified stress model with an added distance adjustment term.
  \item We incorporate the distance adjustment methods into SGD, a state-of-the-art solution method for stress models and confirm its effectiveness.
\end{enumerate}

The rest of this paper is organized as follows:
Section 2 surveys related work on the graph drawing focusing on the stress model;
Sections 3 and 4 presents details of the two proposed methods;
Section 5 explains the computational experiments performed to evaluate the proposed method; and Sections 6 and 7 discuss and conclude this work, respectively.

\section{Related Work}

The origin of graph drawing stress models is traced back to the study of Kamada and Kawai \cite{kamada1989algorithm}, which was an early study of force-directed algorithms \cite{kobourov2013force}.
They proposed an equation equivalent to stress as the energy of a system, in which all node pairs are connected by springs to achieve node proximity and uniformity after placement.
To minimize the energy, they proposed an algorithm that used Newton's method to individually move nodes to find the optimal placement.
However, the energy function is a non-convex function; hence, its optimization is a difficult task.

Gansner et al. \cite{gansner2005graph} focused on the fact that the spring system energy is equivalent to the stress of multidimensional scaling (MDS), which is a dimensionality reduction method. This led them to introduce the majorization technique used in MDS to graph drawing.
Their algorithm is called stress majorization. It monotonically decreases the stress value while simultaneously moving all nodes.
However, a local minimum solution may be reached depending on the initial node arrangement, which makes it difficult to obtain a global minimum solution.
Furthermore, the Kamada--Kawai algorithm and stress majorization require a distance matrix for all node pairs; thus, it takes $O(|V|^3)$ or $O(|V|^2 \log |V| + |V||E|)$ to obtain it.
This method requires a large amount of calculation, and the calculation time problem for graphs with a large number of nodes must be considered.

The sparse stress model proposed by Ortmann et al. \cite{ortmann2016sparse} is an improved solution method for the stress model in terms of computational complexity.
Instead of finding the shortest paths for all node pairs, the shortest paths from some selected pivots are found in the sparse stress model.
The stress on those node pairs is then optimized.
A similar approximation approach for graph drawing was also found in the Pivot MDS proposed by Brandes \cite{brandes2006eigensolver}.

Zheng et al. \cite{zheng2018graph} suggested a method of optimizing the stress function using SGD, which has been widely used in the machine learning field in recent years.
SGD is a gradient descent method variant that uses a random selection algorithm that effectively reduces the stress function value without falling into poor local optimal solutions.
They also proposed SparseSGD, which uses a sparse approximation of the stress function to minimize the stress function without calculating the distance among all node pairs.
Therefore, SGD is currently considered as the most suitable method for minimizing the stress function in terms of both optimization performance and computational complexity.

Accordingly, several other approaches have been proposed to improve the computational complexity of solving stress models.
The maxent stress model proposed by Gansner et al. \cite{gansner2012maxent} improves the computational complexity while deteriorating the drawing quality due to sparse approximation.
In this model, the entropy term using the distance among nodes is added to prevent the nodes from being placed too close to each other due to sparse approximation.
The original idea behind the maxent stress model is found in the binary stress model proposed by Koren and Civril \cite{koren2008binary}.
The maxent stress model was later extended by Meyerhenke et al. \cite{7889042} to a multilevel approach for large graphs.
Gansner et al. further extended their model and proposed COAST \cite{gansner2013coast}, which was reformulated as a convex programming problem.
Another notable approach for improving the computational complexity is the low-rank stress majorization proposed by Khoury et al. \cite{khoury2012drawing}.
They applied singular value decomposition to the graph Laplacian and reduced the computational complexity by using a low-rank Laplacian.
Sublinear-time algorithms for stress minimization were proposed by Meidiana et al. \cite{meidiana2021sublinear}.
They extended the sublinear time force-directed graph drawing through random sampling \cite{meidiana2020sublinear,meidiana2021sublinear} to a stress model.

Incorporating various drawing constraints is one of the main directions for extending stress models.
For hierarchical graph drawing, Dwyer and Koren employed a stress model that added linear constraints regarding the hierarchy \cite{dwyer2005dig}, unlike the Sugiyama framework \cite{Sugiyama1981} that is widely used.
They showed that a stress model with linear constraints can be optimized by iteratively solving a constrained quadratic programming problem using the majorization technique.
They later extended their approach to drawing general graphs with linear constraints \cite{dwyer2006stress,dwyer2006ipsep,dwyer2009constrained}.
They also employed gradient projection to avoid directly solving a constrained quadratic programming problem.
Stress model applications to radial graph drawing \cite{di2013spine} can be found in the papers of Brandes and Pich \cite{brandes2009more} and Xue et al. \cite{xue2022target}.
In their studies, a constraint for concentrically arranging nodes was integrated to the objective function of the stress model.
Stress model extensions to graph drawing in a non-Euclidean space \cite{kobourov2005non} were discussed by Miller et al. \cite{miller2022spherical,miller2022browser}.
They showed that a stress model can be applied to graphs drawn on spherical and hyperboloid planes by calculating the distance among nodes in these spaces.
Meanwhile, Wang et al. surveyed various constrained graph drawings and reorganized them into a unified framework by incorporating them in to a stress model \cite{wang2017revisiting}.

In summary, stress models have been studied since the automatic graph drawing using computers became practical.
During which, numerous improvements to the solution methods and model extensions have been made.
However, the distance matrix in the stress model only uses the graph-theoretic shortest distance, and its influence on the drawing results has hardly been clarified.
We focus herein on that point and clarify the appropriate methods for adjusting the distance matrix and their effects.

\section{Low-rank Approximation of the Distance Matrix}

Graph drawing using stress models is closely related to MDS, which is one of the most popular dimensionality reduction methods.
The stress model formulation is equivalent to metric MDS. Hence, we also consider the approach of applying classical MDS to graph drawing \cite{brandes2006eigensolver}.
First, we will explain how distance matrices are processed in classical MDS.

Let $Y_i = (y_{i1}, y_{i2}, \cdots, y_{i|V|})^T$ denote the coordinates of the $i$th node in a $|V|$-dimensional space.
$Y = (Y_1, Y_2, \cdots, Y_{|V|})^T$ is assumed to accurately reflect the distance matrix $D$ relationship.
Note that $Y_i$ may have non-real components due to the non-Euclidean property of the distance matrix.
The $(i, j)$ element of the distance matrix $D$ represents the distance between $Y_i$ and $Y_j$, while the (i, j) element of the inner product matrix $K$ represents the inner product of $Y_i$ and $Y_j$.
The double centering matrix $H$ is defined as follows:
\begin{equation}
  H = I - \frac{1}{|V|} J
\end{equation}
where $I$ is the identity matrix of order $|V|$, and $J$ is a square matrix of order $|V|$ with all ones.
The inner product matrix $K$ is calculated as follows from the distance matrix $D$:
\begin{equation}
  \label{eq:double-centering}
  K = - \frac{1}{2} H \left(D \bigodot D\right) H
\end{equation}
where $A \bigodot B$ is the element-wise product of matrices $A$ and $B$, which is also known as the Hadmard product of $A$ and $B$, respectively.

Solving the eigenvalue problem yields an eigenvalue $\lambda$ and an eigenvector $u$ that satisfy the following relationship:
\begin{equation}
  K v = \lambda u
\end{equation}
A square matrix of order $n$ has $n$ eigenvalue and eigenvector pairs with duplicates.
Let the $i$th eigenvalue be $\lambda_i$ and eigenvector be $u_i = (u_{i1}, u_{i2}, \cdots, u_{i|V|})^T$.
We assume the eigenvalues to be in a descending order of value, that is, $\lambda_i \geq \lambda_j$ if $i < j$.

In classical MDS, the following relationship with the eigenvalue decomposition is assumed:
\begin{equation}
  X^T X \simeq Y^T Y = K = U \Lambda U^T
\end{equation}
where $\Lambda = \text{diag}(\lambda_1, \lambda_2, \cdots, \lambda_{|V|})$, and $U$ is the square matrix of order $|V|$ with eigenvector $u_i$ corresponding to $\lambda_i$ in the $i$th column.
The $m$-dimensional coordinate X of the node is obtained as follows using the top m eigenvalues and the corresponding eigenvectors:
\begin{equation}
  X_i = \left(\sqrt{\lambda_1} u_{1i}, \sqrt{\lambda_2} u_{2i}, \cdots, \sqrt{\lambda_m} u_{mi} \right)^T
\end{equation}
The inner product matrix has negative eigenvalues if the distance relationship between the nodes cannot be embedded in a Euclidean space.
Negative eigenvalues are usually treated as 0 in classical MDS.

Changing some of the eigenvalue matrix $\Lambda$ elements to 0 means reducing the corresponding dimension.
The number of the non-zero eigenvalues of a matrix is equal to its rank; hence, we consider the replacement operation given some eigenvalues with a 0 low-rank approximation.
We now consider how to obtain the distance matrix from the inner product matrix with some reduced dimensions.
First, we introduce $\delta(\lambda_k)$ to determine whether or not to reduce the dimension corresponding to $\lambda_k$ and rewrite $X_i$ as follows:
\begin{equation}
  \begin{aligned}
    \label{eq:x-definition}
    X_i             & = \left( \delta(\lambda_1) \sqrt{\lambda_1} u_{1i}, \delta(\lambda_2) \sqrt{\lambda_2} u_{2i}, \cdots, \delta(\lambda_{|V|}) \sqrt{\lambda_{|V|}} u_{|V|i} \right)^T \\
    \delta(\lambda) & = \begin{cases}
      0 \quad \text{if the dimension corresponding to } \lambda \text{ is reduced} \\
      1 \quad \text{otherwise}
    \end{cases}
  \end{aligned}
\end{equation}
This $\delta(\lambda)$ definition is abstract. A concrete definition will be introduced later.
Note that $X_i$ here is not intended as the graph drawing coordinates, and it may be a complex vector.

We also consider a inner product matrix $K'$, in which the $(i, j)$ element is the inner product of $X_i$ and $X_j$.
We tread $X_i$ as a complex vector; therefore, $K'$ is represented as $X X^{*}$ using the adjoint matrix $X^{*}$ of $X$.
$K'$ is decomposed as follows into the components of each dimension:
\begin{equation}
  \begin{aligned}
    K'            & = \sum_{k = 1}^{|V|} K'^{(k)}                                                           \\
    K'^{(k)}      & = \left[ K'^{(k)}_{ij} \right]_{|V| \times |V|, (1 \leq i \leq |V|, 1 \leq j \leq |V|)} \\
    K'^{(k)}_{ij} & = \overline{x_{ik}} x_{jk}
  \end{aligned}
\end{equation}
where $\overline{a}$ is the complex conjugate of $a$.
$K'^{(k)}$ is a real symmetric matrix, even though $X_i$ is a complex vector because all elements of the $k$th dimension of $X$ have only real or imaginary parts from Eq. (\ref{eq:x-definition}).

The double centering matrix is not an invertible matrix; thus, it does not have an inverse matrix.
Consequently, the distance matrix cannot be obtained from the inner product matrix using Eq. (\ref{eq:double-centering}).
However, the following distance matrix can be obtained from the inner product matrix using the relationship between the distance and the inner product of the coordinate vectors:
\begin{equation}
  \begin{aligned}
    D'_{ij} & = |X_i - X_j|                                                                                                                                \\
            & = \sqrt{\sum_{k=1}^{|V|} \overline{(x_{ik} - x_{jk})} (x_{ik} - x_{jk})}                                                                     \\
            & = \sqrt{\sum_{k=1}^{|V|} \overline{x_{ik}} x_{ik} - 2 \sum_{k=1}^{|V|} \overline{x_{ik}} x_{jk} + \sum_{k=1}^{|V|} \overline{x_{jk}} x_{jk}} \\
            & = \sqrt{|X_i|^2 - 2 X_j X_i^{*} + |X_j|^2}                                                                                                   \\
            & = \sqrt{K'_{ii} - 2 K'_{ij} + K'_{jj}}
  \end{aligned}
\end{equation}
All $D'$ elements are real numbers, even if $X_i$ is a complex vector because the following relationship holds:
\begin{equation}
  \begin{aligned}
    K'_{ii} - 2 K'_{ij} + K'_{jj} = \sum_{k=1}^{|V|} \overline{(x_{ik} - x_{jk})} (x_{ik} - x_{jk}) \geq 0
  \end{aligned}
\end{equation}
However, $D'$ may contain zeros in non-diagonal components.
This is not suitable for graph drawing; thus, a minimum distance threshold $d_{\text{min}}$ is set for non-diagonal components. That value is set when the distance is less than the threshold.

The following procedure for obtaining a low-rank approximation of the distance matrix by removing some dimensions from the distance matrix is constructed based on the observations:
\begin{enumerate}
  \item Compute the original distance matrix $D$.
  \item Compute the inner product matrix $K$ from the distance matrix $D$.
  \item Compute the eigenvalue decomposition $K = U \Lambda U^T$.
  \item Compute the adjusted distance matrix $D'$ from $K'$.
  \item Solve the stress model using the adjusted distance matrix $D'$.
\end{enumerate}
Several methods for selecting the eigenvalues to be set to 0 are conceivable.
We employed the following method in this work:
\begin{enumerate}
  \item Let $p$ be a real number satisfying $0 \leq p < 100$.
  \item Let $\mu$ be the $p$-percentile of the absolute value of the eigenvalues.
  \item Eigenvalues with an absolute value of less than $\mu$ are set to 0.
\end{enumerate}
We concretely define $\delta(\lambda)$ for computing $K'$ as follows:
\begin{equation}
  \delta(\lambda) = \begin{cases}
    0 \quad |\lambda| < \mu \\
    1 \quad |\lambda| \geq \mu
  \end{cases}
\end{equation}
The result keeps the original distance matrix if $p = 0$.
This criterion leaves some negative eigenvalues unlike classical MDS.
This may be useful in removing the less influential higher-dimensional components while preserving original distance relationships.

In terms of the time computational complexity, the proposed method requires $O(|V|^3)$ for the eigenvalue decomposition in addition to the original graph drawing algorithms using the stress model.
The time computational complexity of the major algorithms using the stress model depends on the distance matrix calculation.
The two major all-pair shortest path algorithms, namely the Floyd--Warshall algorithm for dense graphs and Johnson's algorithm for sparse graphs, require $O(|V|^3)$ and $O(|V|^2 \log |V| + |V||E|)$, respectively.
Adding the eigenvalue decomposition worsens the time computational complexity for sparse graphs, but does not change the time computational complexity for dense ones.

We utilize SGD to solve the stress model.
The SGD method for graph drawing effectively minimizes the stress function while escaping the poor quality local minima.
This method solves the stress model using a low-rank approximation of the distance matrix with SGD and is referred to as the low-rank SGD herein.

\section{Distance-adjusted Stress Model}

Algorithms using the sparse approximation of the distance matrix (e.g., SparseSGD) are superior in terms of the time computational complexity.
LR-SGD requires the $O(|V|^3)$ time computational complexity of the eigenvalue decomposition as a preprocess.
We consider different distance matrix adjustment methods to avoid this preprocessing and improve the time computational complexity.

In a conventional stress model, the distance matrix $D$ is treated as a constant. We modify the model using the adjusted distance matrix $D'$ as a variable.
We add a minimization of their Frobenius norm $|D - D'|_F$ to the stress model to make the adjusted distance matrix reflect the original distance matrix features.
Let $\alpha$ be the weight that satisfies $0 \leq \alpha < 1$. We consider the following weighted sum of the stress function for $D'$ and the Frobenius norm between $D$ and $D'$ as the objective function:
\begin{equation}
  \begin{aligned}
    S\left(X, D'\right) & = \alpha \sum_{i < j} w_{ij} \left(|X_i - X_j| - d'_{ij}\right)^2 + (1 - \alpha) \left|D - D'\right|_F^2                        \\
                        & = \sum_{i < j} \left(\alpha w_{ij} \left(|X_i - X_j| - d'_{ij}\right)^2 + 2 (1 - \alpha) \left(d_{ij} - d'_{ij}\right)^2\right)
  \end{aligned}
\end{equation}
We call this modified model as the distance adjusted stress model.

Let the initial value of $D'$ be $D$.
Optimizing $D'$ adjusts the distance matrix closer to the current $X$.
In this work, we alternately optimize $X$ and $D'$.
Accordingly, the two following phases are used: (1) a phase in which $D'$ is fixed, and $X$ is optimized, and (2) a phase in which $X$ is fixed, and $D'$ is optimized.
Phase (1) is the same as the conventional stress model.
For Phase (2), the stress function is a convex function with respect to $D'$.
Therefore, the partial derivative of the modified stress function with respect to $d'_{ij}$ is set to 0. The optimum $d'_{ij}$ is obtained as follows:
\begin{equation}
  \begin{aligned}
    \frac{\partial}{\partial d'_{ij}} S\left(X, D'\right) & = -2 \left( \alpha w_{ij} |X_i - X_j| + 2 (1 - \alpha) (d_{ij} - d'_{ij}) \right) = 0      \\
    d'_{ij}                                               & = \frac{\alpha w_{ij} |X_i - X_j| + 2 (1 - \alpha) d_{ij}}{\alpha w_{ij} + 2 (1 - \alpha)}
  \end{aligned}
\end{equation}
The resulting optimal $d'_{ij}$ is the $\alpha$-weighted sum of $d_{ij}$ and the current distance between $X_i$ and $X_j$.
A 0 $\alpha$ is equivalent to the conventional stress model.
The larger the $\alpha$, the stronger the power for distance adjustment.
Similar to LR-SGD, we set a minimum distance threshold $d_{\text{min}}$ to stabilize the algorithm and prevent the updated distances from falling below the threshold.
To avoid the influence of far-distant node pairs randomly placed at the beginning of the SGD iteration, the original distance is maintained if the corrected distance exceeds the original distance.
Updating all distances takes as much time computational complexity as the number of elements of required $d'_{ij}$.

We give an engineering interpretation to the update formula for $d'_{ij}$ by assuming that the updated value is expressed as a $c$-weighted sum of $|X_i - X_j|$ and $d_{ij}$:
\begin{equation}
  \frac{\alpha w_{ij} |X_i - X_j| + 2 (1 - \alpha) d_{ij}}{\alpha w_{ij} + 2 (1 - \alpha)} = c |X_i - X_j| + (1 - c) d_{ij}
\end{equation}
Solving this for $\alpha$ yields
\begin{equation}
  \alpha = \frac{1}{1 + r w_{ij}}
\end{equation}
where $r = (1 - c) / 2c$.
A small $w_{ij}$ requires a larger $\alpha$ to achieve the same $c$.
Assuming that $w_{ij} = d_{ij}^{-2}$, two nodes with a shorter ideal distance are more susceptible to the influence of $\alpha$.
The neighboring nodes are susceptible to the distance adjustment. Therefore, the distance relationship in the global structure of the graph is thought to likely be maintained.

The distance-adjusted stress model can be used with minor algorithm changes using the conventional stress model.
This modification does not change the time computational complexity of the original algorithm. It can also be processed with little additional computation time.
We solve the distance-adjusted stress model by SGD, as in the previous section.
Algorithm \ref{algorithm:daf-sgd} shows the pseudo-code of FullSGD with a distance adjustment.

\begin{algorithm}[t]
  \SetKwInOut{Input}{input}
  \Input{graph $G = (V, E)$}
  \Input{distane adjustment parameter $\alpha$}
  \Input{minimum distance threshold $d_{\min}$}
  $D \leftarrow \text{AllPairsShortestPaths}(G)$\;
  $D' \leftarrow \text{copy}(D)$\;
  $X \leftarrow \text{InitialPlacemnt}(G)$\;
  $\text{node\_pairs} \leftarrow \{(i, j) \mid i, j \in V, i < j\}$\;
  \For{$t \leftarrow 0$ \KwTo $\text{number\_of\_itrations} - 1$}{
  $\text{shuffle}(\text{node\_pairs})$\;
  \tcc{node placement phase}
  \ForEach{$(i,j) \in \text{node\_pairs}$}{
  $\mu \leftarrow \min(1, w_{ij} \eta(t))$\;
  $r \leftarrow \frac{|X_i - X_j - d'_{ij}}{2} \frac{X_i - X_j}{|X_i - X_j|}$\;
  $X_i \leftarrow X_i - \mu r$\;
  $X_j \leftarrow X_j + \mu r$\;
  }
  \tcc{distance adjustment phase}
  \ForEach{$(i,j) \in \text{node\_pairs}$}{
  $d'_{ij} \leftarrow \text{clamp}\left( d_{\min}, \frac{\alpha w_{ij} |X_i - X_j| + 2 (1 - \alpha) d_{ij}}{\alpha w_{ij} + 2 ( 1 - \alpha)}, d_{ij} \right)$\;
  }
  }
  \Return X\;
  \caption{Pseudo-code for the Ditance-adjusted FullSGD (DAF-SGD).}
  \label{algorithm:daf-sgd}
\end{algorithm}

\begin{algorithm}[p]
  \SetKwInOut{Input}{input}
  \Input{graph $G = (V, E)$}
  \Input{distane adjustment parameter $\alpha$}
  \Input{minimum distance threshold $d_{\min}$}
  \Input{number of pivots $h$}
  $P \leftarrow \text{ChoosePivots(G, h)}$\;
  $D \leftarrow \text{SparseShortestPaths}(G, P)$\;
  $D' \leftarrow \text{copy}(D)$\;
  $w_{pi} \leftarrow 0 \forall p \in P, i \in V$\;
  \ForEach{$(p, i) \in P \times V, p \notin N(i)$}{
  $s \leftarrow |\{j \in R(p) \mid d_{pj} \leq d_{pi} / 2\}|$\;
  $w'_{ip} \leftarrow s w_{ip}$\;
  }
  \ForEach{$(i, j) \in E $}{
  $w'_{ij} \leftarrow w_{ij}$\;
  $w'_{ji} \leftarrow w_{ji}$\;
  }
  $X \leftarrow \text{InitialPlacemnt}(G)$\;
  $\text{node\_pairs} \leftarrow E \cup (V \times P)$\;
  \For{$t \leftarrow 0$ \KwTo $\text{number\_of\_itrations} - 1$}{
  $\text{shuffle}(\text{node\_pairs})$\;
  \tcc{node placement phase}
  \ForEach{$(i,j) \in \text{node\_pairs}$}{
  $\mu_i \leftarrow \min(1, w'_{ij} \eta(t))$\;
  $\mu_j \leftarrow \min(1, w'_{ji} \eta(t))$\;
  $r \leftarrow \frac{|X_i - X_j - d'_{ij}}{2} \frac{X_i - X_j}{|X_i - X_j|}$\;
  $X_i \leftarrow X_i - \mu_i r$\;
  $X_j \leftarrow X_j + \mu_j r$\;
  }
  \tcc{distance adjustment phase}
  \ForEach{$(i,j) \in \text{node\_pairs}$}{
  $d'_{ij} \leftarrow \text{clamp}\left( d_{\min}, \frac{\alpha w_{ij} |X_i - X_j| + 2 (1 - \alpha) d_{ij}}{\alpha w_{ij} + 2 ( 1 - \alpha)}, d_{ij} \right)$\;
  }
  }
  \Return X\;
  \caption{Pseudo-code for the Ditance-djusted SparseSGD (DAS-SGD).}
  \label{algorithm:das-sgd}
\end{algorithm}

The distance-adjusted stress model works best in sparse stress models because it only adjusts the necessary distances.
The sparse stress model is a modification of the stress function that computes only the distance for some node pairs instead of computing the distances for all node pairs.
We apply the distance-adjusted stress model to SparseSGD, which is a sparse SGD approximation.
Algorithm \ref{algorithm:das-sgd} shows the pseudo-code of SparseSGD with a distance adjustment.

In contrast to SparseSGD, the SGD that computes for all node pairs is called FullSGD.
We denote the distance-adjusted FullSGD and SparseSGD as DAF-SGD and DAS-SGD, respectively, and collectively refer to these as DA-SGD.

\section{Computational Experiments}

\subsection{Experimental Conditions}

We conducted computational experiments to evaluate the proposed LR-SGD and DA-SGD algorithms from several viewpoints.
Various quality metrics have been proposed to assess graph drawing results. Stress is one of them.
From the viewpoint of the drawing result evaluation, stress must be measured using the original distance matrix instead of the adjusted distance matrix.
Solving the stress model with the adjusted distance matrix worsens the stress on the original distance matrix.

Prior to the experiment, we generated some drawing results using the proposed algorithms, confirming the existence of drawing results with a subjectively good visibility despite stress worsening.
We reasoned that aside from stress, the other quality metrics were improved with the distance adjustment.
Therefore, in addition to stress, we used the eight quality metrics adopted by Ahmed et al. \cite{ahmed2022multicriteria} to characterize the drawing results of the proposed algorithms.
For the quality metrics listed below, higher values are desirable for the aspect ratio and neighborhood preservation, while lower values are desirable for the quality metrics.

\paragraph{Ideal Edge Lengths}
The ideal edge length formulated as follows is defined as the sum of the squares of the relative error between the distance and the ideal distance for the pairs of vertices (not all vertex pairs) connected by edges on the graph.
\begin{equation}
  f(X)_{\text{IL}} = \sum_{(i, j) \in E} \left( \frac{|X_i - X_j| - d_{ij}}{d_{ij}} \right)^2
\end{equation}
Although this is part of the stress function, this is useful for evaluating the distance relationships only where edges exist.

\paragraph{Neighborhood Preservation}
Neighborhood preservation evaluates the extent to which the adjacency relationships on the graph are reflected in the drawing result proximity.
The Jacquard distance between the edge set of the shape graph $G_N = (V_N, E_N)$ constructed from the drawing result coordinates \cite{eades2017shape,nguyen2017proxy} and the edge set of the original graph is used as a quality metrics.
\begin{equation}
  f(X)_{\text{NP}} = \frac{|E \cap E_N|}{|E \cup E_N|}
\end{equation}
We utilized a $k$-nearest neighbor graph with the same number of edges as its degree from each node by referring to the paper of Ahmed et al. \cite{ahmed2022multicriteria}.

\paragraph{Crossing Number}
Minimizing the number of edge crossings is widely accepted as a classical aesthetic criterion for graph drawing.
We also used this as a quality metric of the drawing results.

\paragraph{Crossing Angle}
Edge crossings are often unavoidable in real-world graphs.
When edge crossings occur, it is desirable that the angles they form are right angles.
We followed previous research \cite{eades2010force,ahmed2022multicriteria} and formulated the crossing angle as the squared sum of the cosines of the angles formed by the crossing edges:
\begin{equation}
  f(X)_{\text{CA}} = \sum_{((i, j), (k, l)) \in C} \left( \frac{(X_i - X_j) (X_k - X_l)^T}{\left| X_i - X_j \right| \left| X_k - X_l \right| }\right)^2
\end{equation}
where $C \subseteq E \times E$ is the set of crossing edges.

\paragraph{Aspect Ratio}
The aspect ratio herein is the ratio of the height to the width of the drawing result area.
The worst aspect ratio considering the drawing area rotation is obtained by taking the singular value of the centered $X$ \cite{ahmed2022multicriteria}.
The centered X formally has two nonzero singular values $\sigma_1, \sigma_2 (\sigma_1 > \sigma_2)$. The aspect ratio is denoted by $\sigma_2 / \sigma_1$.

\paragraph{Angular Resolution}
Let $\phi_{ijk}$ be the angle formed by the two edges of $(i, j)$ and $(j, k)$ that share an endpoint.
Although the angular resolution is sometimes defined as the minimum $\phi_{ijk}$, we used the following angular energy formulation \cite{ahmed2022multicriteria} to evaluate the angular resolution of entire drawing result:
\begin{equation}
  f(X)_{\text{ANR}} = \sum_{(i, j), (j, k) \in E} \exp(-\phi_{ijk})
\end{equation}

\paragraph{Node Resolution}
The node resolution evaluates the node placement $X$ uniformity.
We employed the following loss function \cite{ahmed2022multicriteria}:
\begin{equation}
  f(X)_{\text{NR}} = \sum_{i < j} \left(1 - \frac{|X_i - X_j|}{r d_{\max}} \right)^2
\end{equation}
where $d_{\max} = \max_{i < j}|X_i - X_j|$ and $r = 1 / \sqrt{|V|}$.

\paragraph{Gabriel Graph Property}
The Gabriel graph is a graph in which the edges are connected so that no other nodes are included in a circle whose diameter is a line segment connecting nodes $i$ and $j$ ($i, j \in V$).
The Gabriel graph property evaluates how close the drawing result is to the Gabriel graph.
The calculations are performed such that a penalty is applied when a node is included in a circle whose diameter is the graph edge:
\begin{equation}
  f(X)_{\text{GB}} = \sum_{(i, j) \in E} \sum_{\substack{k \in V}} \max\left( 0, r_{ij} - |X_k - c_{ij}|\right)^2
\end{equation}
where $c_{ij} = (X_i + X_j) / 2$, and $r_{ij} = |X_i - X_j| / 2$.

The graph data were obtained from the SuiteSparse Matrix Collection \cite{davis2011university} widely used as a graph drawing benchmark.
All graphs were treated as unweighted graphs with edge length 1.

We set the target percentile for the low-rank approximation of the distance matrix as $\{0, 10, \cdots, 90\}$.
Let $\alpha = 1 - 0.5^{k}$ for DAF-SGD and DAS-SGD. We set $k$ as $\{0, 1, \cdots, 9\}$.
The minimum distance threshold $d_{\text{min}}$ was set to 0.1, which was 1/2 of the unit edge length.

The SGD parameters were set to 15 for the number of iterations and 0.1 for $\epsilon$.
The number of pivots for SparseSGD was 200.
SGD is affected by random numbers; hence, we performed 100 trials with different random number seeds for each condition.

\subsection{Results}

\begin{figure*}[p]
  \centering
  \begin{tabular}{c}
    \begin{minipage}[t]{\hsize}
      \centering
      \includegraphics[width=\hsize]{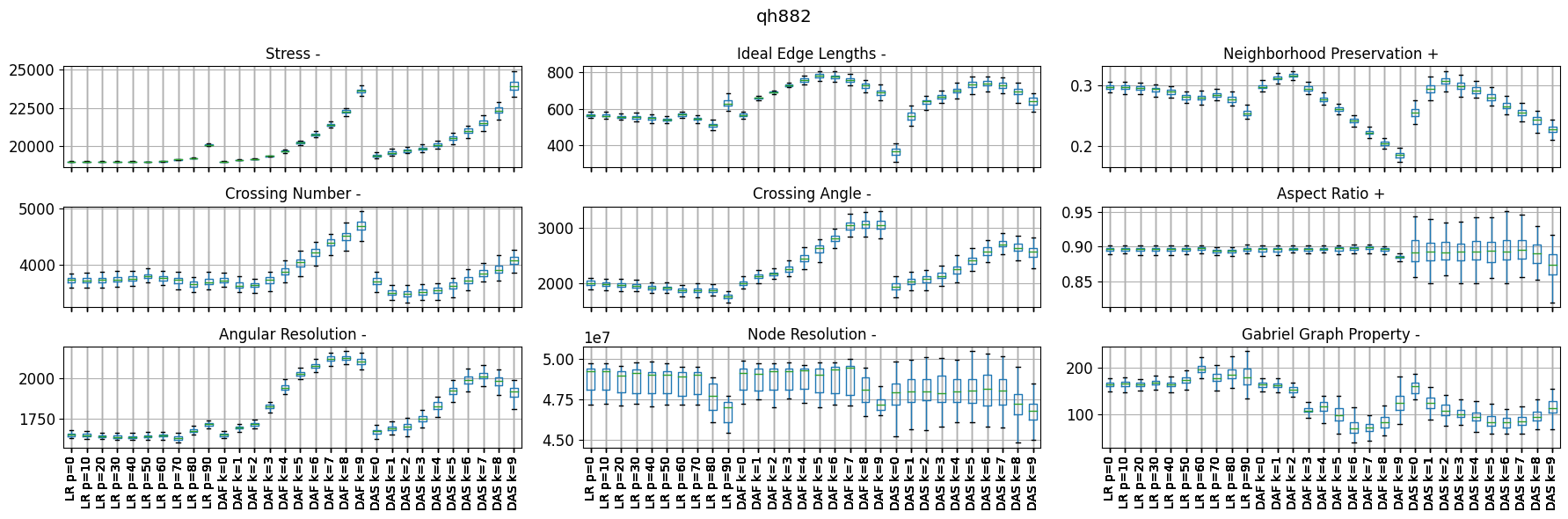}
    \end{minipage} \\
    \begin{minipage}[t]{\hsize}
      \centering
      \includegraphics[width=\hsize]{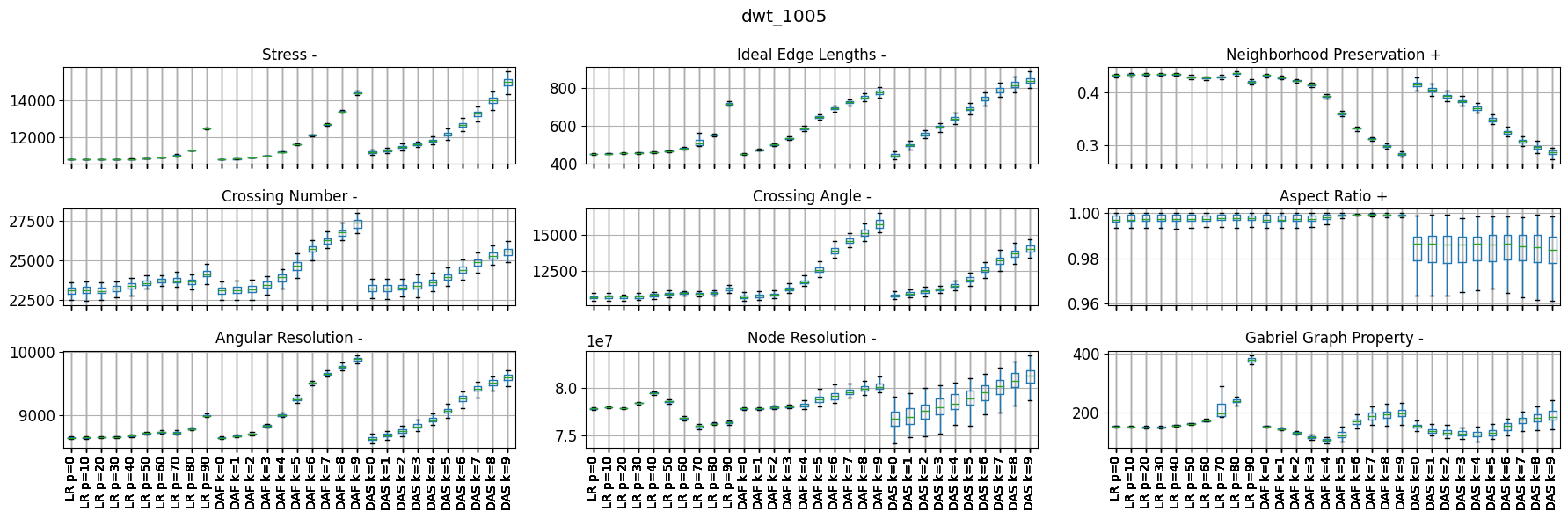}
    \end{minipage} \\
    \begin{minipage}[t]{\hsize}
      \centering
      \includegraphics[width=\hsize]{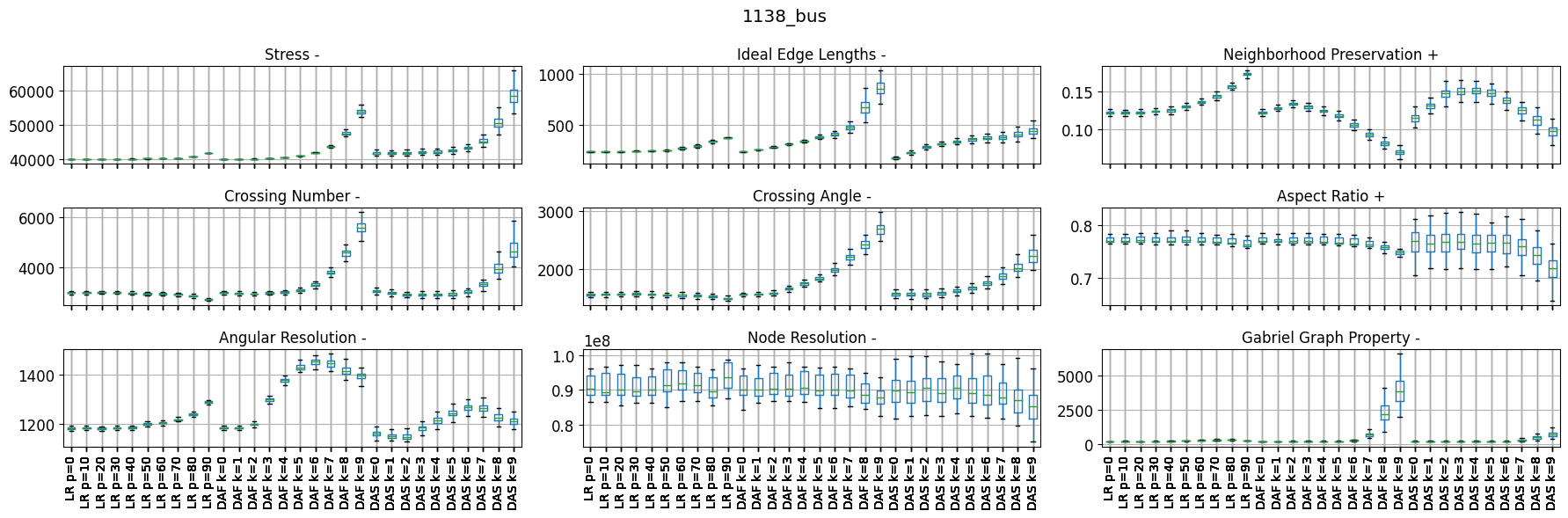}
    \end{minipage} \\
    \begin{minipage}[t]{\hsize}
      \centering
      \includegraphics[width=\hsize]{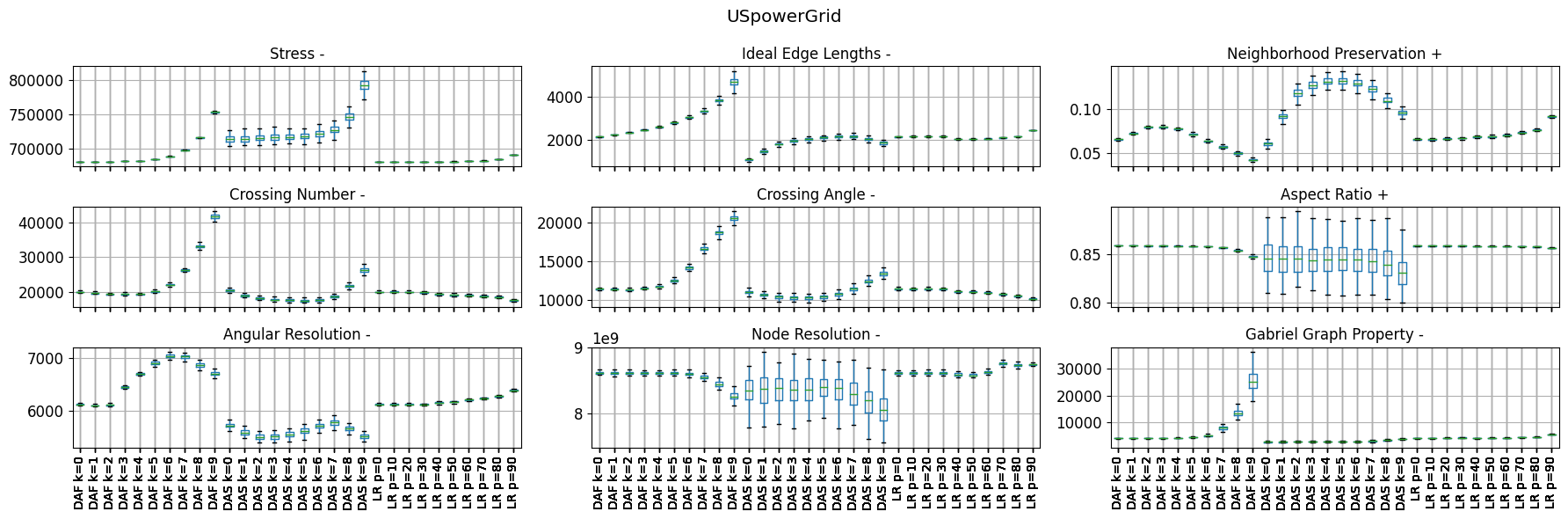}
    \end{minipage}
  \end{tabular}
  \vspace{-0.5cm}
  \caption{Box plots showing th change in the quality metrics caused by the distance adjustment in the qh882, dwt\_1005, 1138\_bus, and USpowerGrid graphs.}
  \label{fig:boxplots}
\end{figure*}

\begin{figure*}[p]
  \begin{tabular}{c}
    \begin{minipage}[t]{\hsize}
      \centering
      \includegraphics[width=\hsize]{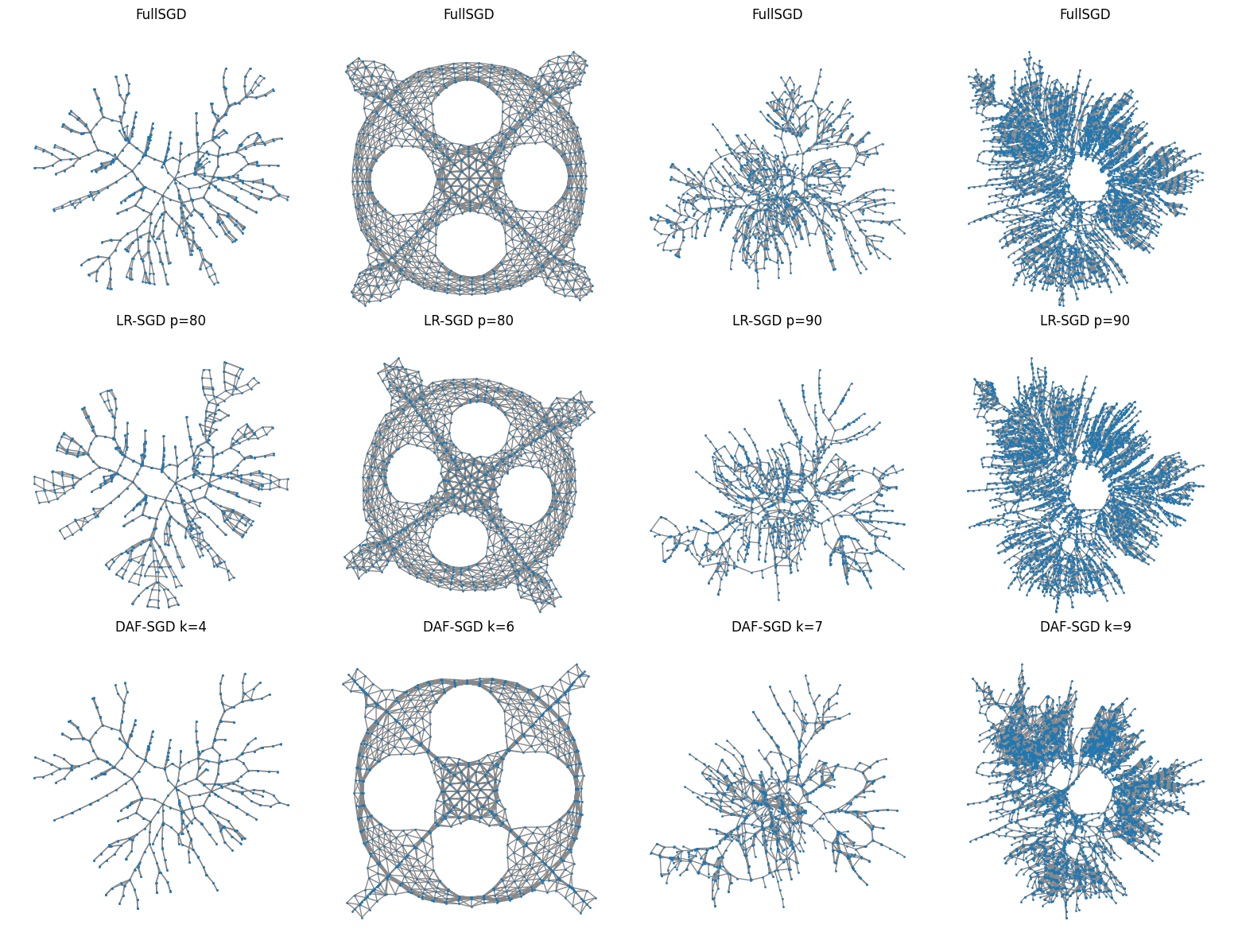}
      \caption{Examples of the drawing results for the qh882, dwt\_1005, 1138\_bus, and USpowerGrid graphs using FullSGD, LR-SGD and DAF-SGD.}
      \label{fig:drawing_result_full}
    \end{minipage} \\
    \begin{minipage}[t]{\hsize}
      \centering
      \includegraphics[width=\hsize]{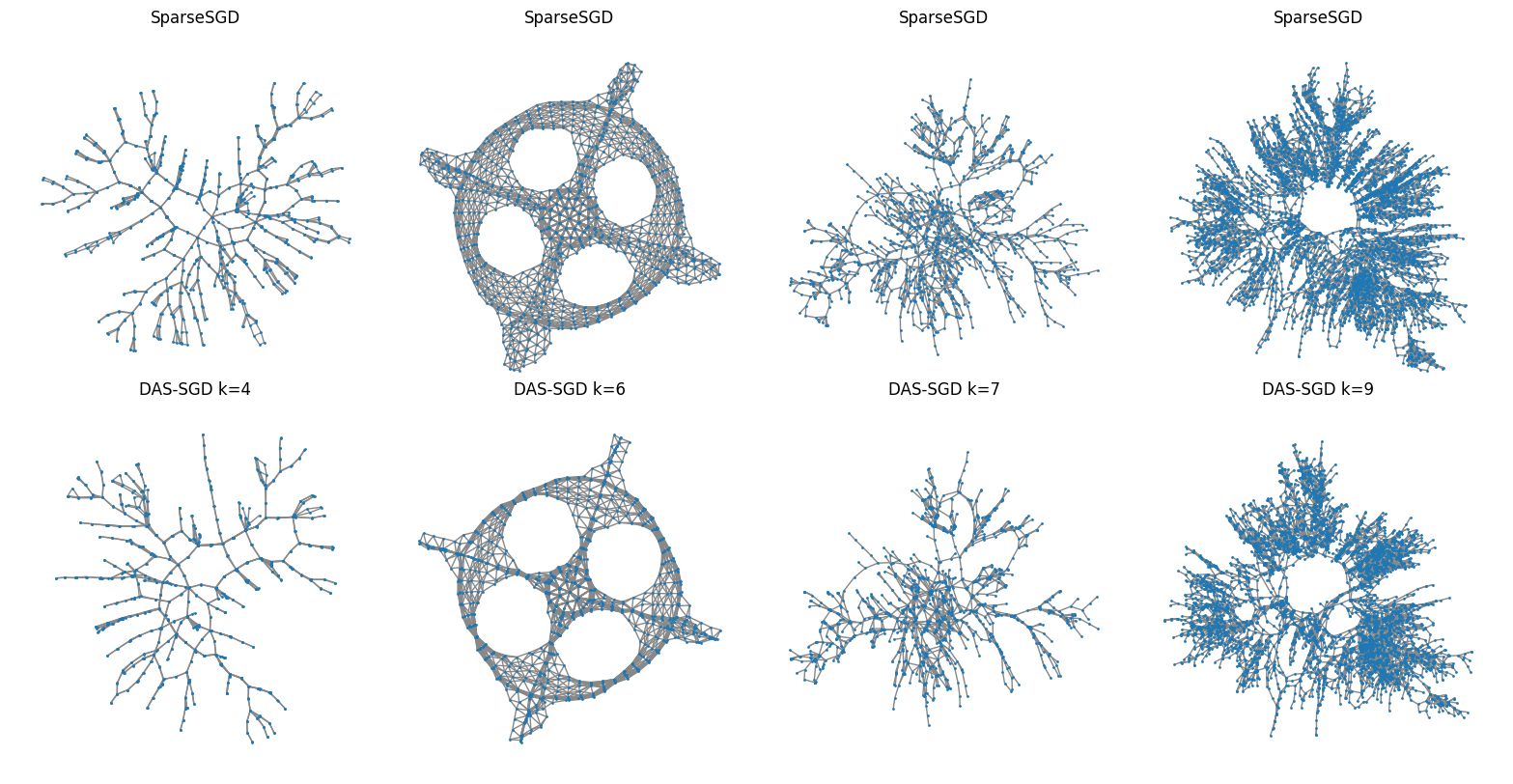}
      \caption{Examples of the drawing results for the qh882, dwt\_1005, 1138\_bus, and USpowerGrid graphs using SparseSGD and DAS-SGD.}
      \label{fig:drawing_result_sparse}
    \end{minipage}
  \end{tabular}
\end{figure*}

We first observed the behavior of the proposed method on four graphs, namely qh882 ($|V| = 882, |E| = 1533$), dwt\_1005 ($|V| = 1005, |E| = 3808$), 1138\_bus ($|V| = 1138, |E| = 1458$), and USpowerGrid ($|V| = 4941, |E| = 6594$).
Figure \ref{fig:boxplots} depicts boxplots showing the distribution of the nine quality metrics of the drawing results under the conditions of each graph.
The quality metrics confirmed to be improved under some conditions were the ideal edge length, node resolution, neighborhood preservation, and Gabriel graph property.
The other quality metrics tended to worsen as the degree of the distance adjustment increased.
We confirmed that stress worsened as the degree of distance adjustment increased.
It was calculated using the original distance; therefore, it was not compatible with the distance changed using the proposed method.
We found a slight difference in the distribution trends between DAF-SGD and DAS-SGD.
Some of these differences may have been caused by the sparse approximations.
In conclusion, the proposed distance-adjusted stress model works in both FullSGD and SparseSGD.
A noteworthy result of DA-SGD was that the Gabriel graph property improved with the moderate distance adjustment of approximately $k = 5$ in qh882 and dwt\_1005.
In qh882, dwt\_1005, and USpowerGrid, the node resolution was also slightly improved at a high $k$ of approximately $k = 8$.
In qh882, 1138\_bus, and USpowerGrid, the neighborhood preservation was improved with a small or medium $k$.
LR-SGD showed a tendency for node resolution improvement with high $p$ in qh882 and dwt\_1005.
In 1138\_bus and USpowerGrid, the neighborhood preservation was improved with a high $p$.

We will now show examples of the drawing results using the three proposed algorithms with the four graphs.
Figure \ref{fig:drawing_result_full} and Figure \ref{fig:drawing_result_sparse} display the drawing results of LR-SGD, DAF-SGD, and DAS-SGD, respectively.
The leftmost columns of all figures were equivalent to the results without a distance adjustment.
The further to the right column, the higher the degree of distance adjustment.
The trends in the drawing results differed between LR-SGD and DA-SGD.
We found no significant difference in the trends of the drawing results between DAF-SGD and DAS-SGD.
These results were consistent with the comparison results using the quality metrics.

\begin{figure}[h]
  \centering
  \includegraphics[width=8cm]{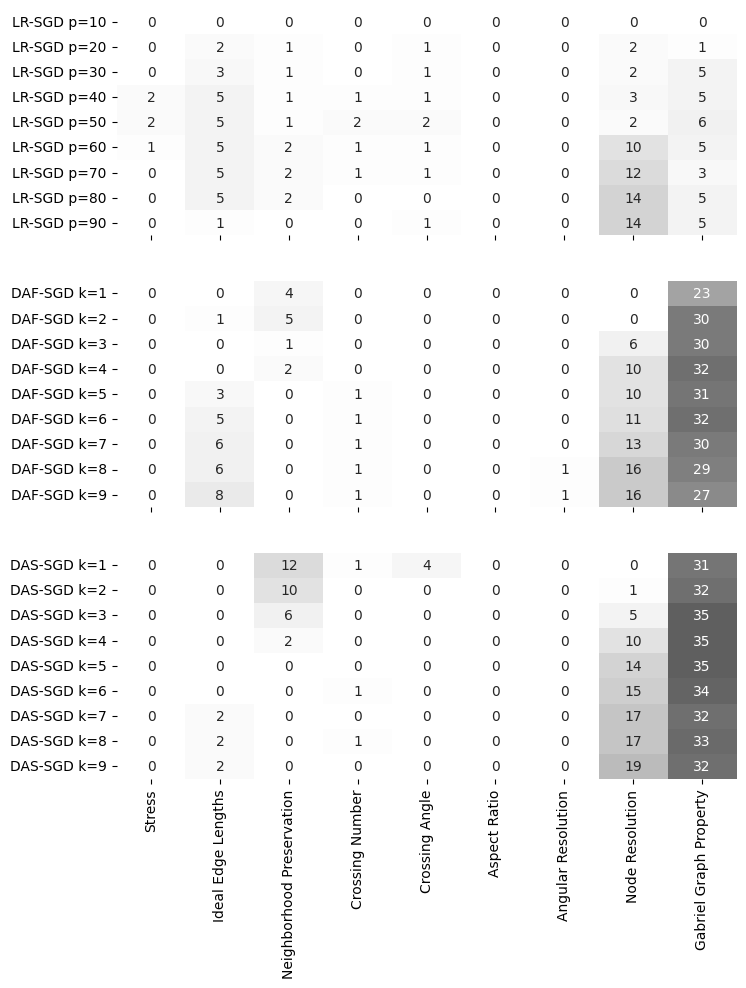}
  \caption{Heatmap of the number of graphs showing a quality metric improvement of 10\% or more.}
  \label{fig:stats}
\end{figure}

Finally, we checked whether or not the properties of the proposed method we confirmed so far can be seen in graphs other than the four graphs mentioned above.
We randomly collected 50 graphs with less than 1000 nodes from the SuiteSparse Matrix Collection.
The collected graphs are as follows: Ecoli\_10NN, G11, G15, G21, G44, G7, G8, G9, M10PI\_n, Trefethen\_20b, USAir97, Vehicle\_10NN, ash292, bcspwr01, bcsstk06, bcsstk19, bcsstm07, breasttissue\_10NN, cage4, cage6, can\_144, can\_445, can\_61, cdde1, cdde6, dermatology\_5NN, dwt\_209, dwt\_66, dwt\_878, dwt\_918, ex1, ex21, hangGlider\_1, iris\_dataset\_30NN, lowThrust\_1, lshp\_265, lshp\_406, lshp\_577, mesh2e1, mesh2em5, micromass\_10NN, mycielskian5, mycielskian8, plat362, spaceShuttleEntry\_1, steam2, tumorAntiAngiogenesis\_2, tumorAntiAngiogenesis\_4, tumorAntiAngiogenesis\_6, and umistfacesnorm\_10NN.
Figure \ref{fig:stats} depicts a heatmap of the number of graphs showing a quality metrics improvement of 10\% or more under each condition.
The most effective improvement was found in the Gabriel graph property using DA-SGD. It demonstrated a quality metrics improvement in approximately 30 out of 50 graphs.
For both LR-SGD and DA-SGD, the rate at which the node resolution was improved when the degree of distance adjustment was high was approximately 20 out of 50.
The improvements in the ideal edge length and the neighborhood preservation were observed in a small number of cases, as well.
Almost no improvement was seen in the other quality metrics.
These results were consistent with those obtained for the four graphs above.

\section{Discussion}

The drawing results in both LR-SGD and DA-SGD changed according to the changes in the degree of distance adjustment.
In LR-SGD, although no significant improvement was observed in the quality metrics for some graphs, we found visual characteristic changes in the drawing results.
Both methods tended to simplify the graph's local structure and emphasize the global structure.
A similar visual effect was also seen in the drawing results of Pivot MDS proposed by Brandes et al. \cite{brandes2006eigensolver}.
They mentioned the application of a spring embedder \cite{Eades1984} after Pivot MDS to improve the subjective drawing quality.
The difference in our proposed method is that the degree of the distance adjustment can be changed as parameters instead of performing a post-processing.
However, in some differences, simplification occurred between the low-rank approximation and the modified stress model.
The manner by which the effect is influenced also depended on the graph structure.
Further theoretical elucidation is needed to understand how the distance matrix adjustment affects the drawing structure.

$(SGD)^2$ proposed by Ahmed et al. \cite{ahmed2022multicriteria} directly optimizes the quality metrics; however, they pointed out that optimizing the crossing number and the Gabriel graph property was difficult.
The distance-adjusted stress model we proposed herein succeeded in improving the Gabriel graph property when compared to a conventional stress model.
The Gabriel graph property is thought to be easily influenced by the local graph structure. Our approach of changing that structure is effective.

\section{Conclusion}

In this study, we focused on the distance matrix used in the stress model for graph drawing and proposed the LR-SGD and DA-SGD methods to draw graphs while adjusting the distance matrix.
LR-SGD performed an eigenvalue decomposition of the distance matrix as a preprocessing for SGD and improved the drawing results by removing the extra space dimensions, in which the node was placed.
DA-SGD added a distance adjustment term to the conventional stress model and solved it using SGD.
We employed SGD here due to its high optimization performance. We, however, believe that the proposed theory can be applied to many existing algorithms.
We also quantitatively evaluated the effect of the proposed method on the drawing results through computer experiments using some benchmark graph data.
The results confirmed that adjusting the distance matrix will cause visual changes in the drawing results and quality metrics changes.

An interesting fact in this study is that adjusting the distance matrix in the stress model affects the quality metrics of the drawing results.
The graph-theoretical distance matrix is an optimization parameter widely used in graph drawing since the introduction of the Kamada--Kawai algorithm.
We showed herein that there is much room for research in terms of providing appropriate ideal distances in stress models, and that they can be integrated into existing research.
One possible direction for future research is the consideration of how to construct appropriate and ideal distances.
Another future challenge is the integration of the proposed approach into the many existing stress model extensions.
We cannot say that the changes in the visual characteristics caused by the proposed method are necessarily reflected in the quality metrics.
Therefore, future tasks must consider quality metrics other than those used in this study and evaluate the human subjective preferences through a user study.

\bibliographystyle{abbrv-doi}
\bibliography{document}
\end{document}